# Spatio-temporal cooperative control Method of Highway Ramp Merge Based on Vehicle-road Coordination

Xiaoxue Xu
*Key Laboratory for Special Area Highway Engineering of Ministry of Education,Chang'an University*
*Xi'an,China*
email :2112534668@qq.com

Maokai Lai
Key Laboratory for Special Area Highway Engineering of Ministry of Education,Chang'an University
Xi'an,China
email :3412303785@qq.com

*HaitaoZhang*
*China Railway Seventh Bureau Group Third Engineering Co., Ltd*
*Xi'an,China*
*email:1657839062@qq.com*

XiangDong
Key Laboratory for Special Area Highway Engineering of Ministry of Education,Chang'an University
Xi'an,China
email:1290740787@qq.come

TaoLi
*Key Laboratory for Special Area Highway Engineering of Ministry of Education,Chang'an University*
email:1186166770@qq.com

JieWu
*Key Laboratory for Special Area Highway Engineering of Ministry of Education,Chang'an University*
*Xi'an,China*
email:2211049290@qq.com

YuanLi
*Key Laboratory for Special Area Highway Engineering of Ministry of Education,Chang'an University*
*Xi'an,China*
*email:157708198@qq.com*

Ting Peng*
*Key Laboratory for Special Area Highway Engineering of Ministry of Education,Chang'an University*
*Xi'an,China*
*email :t.peng@ieee.org*
**Corresponding author*

***Abstract*—The merging area of highway ramps faces multiple challenges, including traffic congestion, collision risks, speed mismatches, driver behavior uncertainties, limited visibility, and bottleneck effects. However, autonomous vehicles engaging in depth coordination between vehicle and road in merging zones, by pre-planning and uploading travel trajectories, can significantly enhance the safety and efficiency of merging zones.In this paper,we mainly introduce mainline priority cooperation method to achieve the time and space cooperative control of highway merge.Vehicle-mounted intelligent units share real-time vehicle status and driving intentions with Road Section Management Units, which pre-plan the spatiotemporal trajectories of vehicle travel. After receiving these trajectories, Vehicle Intelligent Units strictly adhere to them. Through this deep collaboration between vehicles and roads, conflicts in time and space during vehicle travel are eliminated in advance.**

## I. Introduction

Research on cooperative control strategies in ramp merging areas is a hot topic in the fields of intelligent transportation systems and autonomous driving technology. With the development of vehicle-to-everything (V2X) communication technology and the proliferation of intelligent vehicles, many scholars have investigated cooperative merging strategies for intelligent connected and autonomous vehicles (ICAVs), aiming to enhance traffic efficiency and safety by optimizing vehicle trajectories and merging sequences. In terms of optimization control strategies, most literature adopts optimal control strategies such as Mixed-Integer Nonlinear Programming (MINLP)[1-3] pseudospectral methods[4], game theory[5],hierarchical control strategies[6-11],distributed control[12-13], and centralized control[14-15]to precisely plan vehicle merging behaviors and trajectories.

In terms of cooperative control, literature commonly utilizes cooperative adaptive cruise control [16] and multi-agent systems to coordinate the merging behaviors of multiple vehicles, thus improving overall traffic flow. Through vehicle-to-everything (V2X) communication technology, real-time information sharing and collaborative decision-making are achieved among vehicles and between vehicles and infrastructure (V2I) and vehicles (V2V). In terms of simulation validation, all methods are evaluated for their effectiveness using different simulation platforms such as SUMO.Through the aforementioned optimization control strategies, cooperative control methods, and simulation validation approaches, these studies demonstrate effective means to enhance vehicle merging efficiency and safety in complex traffic environments.

Current manual and automated driving technologies still struggle to accurately and promptly acquire the status and driving intentions of surrounding vehicles. Despite efforts to maintain appropriate safety distances between vehicles on highway merge zones, a significant number of traffic accidents still occur. Due to the necessity of maintaining safe distances between vehicles, the capacity for road traffic cannot be further increased, posing a formidable barrier to overcoming traffic

Published in: 2024 12th International Conference on Traffic and Logistic Engineering (ICTLE)
DOI：10.1109/ICTLE62418.2024.10703891

congestion.In order to solve these problems and improve the road traffic efficiency under the premise of ensuring safety, this paper proposes a technical scheme to realize spatial and temporal cooperation by using the road section management unit to prepare the vehicle's space-time trajectory in advance and let the on-board intelligent unit execute it, so as to eliminate the conflict between vehicles in advance.Based on safety distance, collision risk and collision acceleration, this paper presents a spatial-temporal coordinated control strategy for vehicles in ramp confluence area

## II. Main Research Content

On the basis of fully understanding the vehicle inbound model of expressway on-ramp at home and abroad, using the wireless communication and data acquisition technology under the vehicle-road depth cooperation environment, real-time vehicle location and speed information is obtained, and the spatial-temporal collaborative control method of vehicle inbound on expressway on-ramp under the vehicle-road depth cooperation environment is proposed. Finally, SUMO software and its development interface are used to simulate and evaluate the method. The main research contents of this paper are as follows:

- Popose new Road Section Management Units and Vehicle Intelligent Units, and provide detailed descriptions of their new functionalities. Compare and analyze existing communication technologies to determine the sharing method used by the information sharing platform. Based on this, construct a comprehensive and systematic framework for information sharing mechanisms and design interfaces for data presentation by Vehicle Intelligent Units and Road Section Management Units.
- Analyze the technical performance indicators of the real-time information sharing mechanism and the content of data sharing. Conduct detailed research on each module involved in implementing the information sharing mechanism, including the deployment methods of Road Section Management Units, the mechanism of real-time information sharing between vehicles, and the specific methods for positioning Vehicle Intelligent Units on the road.
- Vehicle-mounted intelligent units share real-time vehicle status and driving intentions with Road Section Management Units, which pre-plan the spatiotemporal trajectories of vehicle travel. After receiving these trajectories, Vehicle Intelligent Units strictly adhere to them. Through this deep collaboration between vehicles and roads, conflicts in time and space during vehicle travel are eliminated in advance.
- Study the calculation method of safe distance under spatiotemporal conditions based on differences in vehicle speed, vehicle positioning error, and clock error.
- Propose a collaborative control method to pre-plan vehicle travel trajectories and improve the safety and traffic efficiency of vehicles in highway ramp merging areas.

## III. Quantitative Evaluation of Vehicle Conflict Risk

Given the insufficient research in academia on assessing the severity of collision hazards, this paper selects safety distance and conflict urgency as indicators of conflict risk. This selection is based on a series of theoretical foundations and analytical processes, as outlined below:

### A. *Safety Distance*

Safety distance refers to the minimum distance that should be maintained to prevent collisions between vehicles. When the distance between vehicles is less than this safety distance, the risk of collision significantly increases. Therefore, safety distance is an important indicator for assessing conflict risk.

### B. *Conflict Urgency*

Collision acceleration reflects the danger of vehicle collisions. When the acceleration is low, it only affects passenger comfort. However, when the acceleration is particularly high, it can directly pose lethal problems to passengers. Therefore, collision acceleration is an important indicator for assessing the danger of conflict occurrence.

Urgent acceleration reflects the urgency of collision occurrence. When the urgent acceleration is high, it indicates that the two vehicles are about to collide.

Neither collision acceleration nor urgent acceleration alone can comprehensively reflect the risk of vehicle collisions. Thus, by multiplying them together, the degree of conflict urgency is obtained. This combined metric can simultaneously reflect the danger and urgency of vehicle collisions. Therefore, conflict urgency is considered an important indicator for quantifying conflict risk.

### C. *Deep Cooperative Safety Distance*

The necessity and In both traditional manual driving and autonomous driving, the driving intent of the preceding or surrounding vehicles is often unknown. Therefore, it's necessary to maintain a significant safety distance from the preceding vehicle to ensure timely braking in case of danger and to avoid collisions. However, in the context of this paper, autonomous driving eliminates the possibility of conflicts before they arise by having the road management unit preemptively plan and upload the future trajectories of vehicles. Since the vehicle's pre-planned trajectory is uploaded in advance, the transmission delay of several tens of milliseconds has no effect on the vehicle's future operation. Therefore, transmission delay does not need to be considered. Vehicles share their future trajectories with each other. In this scenario, all vehicles are aware of the driving intent of the preceding vehicle. When the preceding vehicle suddenly decelerates or stops, the following vehicles also decelerate accordingly, preventing collisions from occurring. Hence, there's no need to maintain a large safety distance.

The safety distance includes three components: the safe distance to be maintained between merging vehicles and mainline vehicles when vehicles enter the mainline, the positioning error of the Global Positioning System (GPS), and the timing synchronization error between autonomous driving vehicles and the national time synchronization center.

## IV. The Spatiotemporal Coordinated Control of Vehicles in the Merging Area of Ramp

### A. *The Necessity of Cooperative Control*

The merging zone of highway ramps presents greater safety hazards and lower traffic efficiency compared to the basic sections of highways. Since highways are fully enclosed roadways, drivers tend to speed up when there are fewer vehicles, significantly increasing the accident rate. When traffic volume is high, especially during peak hours, the density of vehicles in the merging zone increases, leading to traffic congestion and raising the likelihood of accidents. In the merging zone, vehicles from the ramp need to change lanes. When these vehicles enter the main lanes, they often have to accelerate from lower speeds to the highway's travel speed. During this process, there is a speed difference between ramp vehicles and those already in the main lanes, resulting in a relatively high collision risk and a significant increase in the possibility of traffic accidents. In China, the speed limit for ramp vehicles on highways is set at 40 km/h, but drivers frequently exceed this limit during actual driving. Scholars studying vehicle speed operations both domestically and internationally have reached a conclusion: when vehicles exceed the prescribed speed limit, the likelihood of safe vehicle operation under road conditions decreases significantly. However, autonomous vehicles engaging in depth coordination between vehicle and road in merging zones, by pre-planning and uploading travel trajectories, can significantly enhance the safety and efficiency of merging zones.

#### *1) Highway ramp merging zone safety analysis:*

The highway ramp merging area requires vehicles to change lanes in order to merge onto the main road. The key factor in lane changing is to find a suitable gap on the main road. Vehicles entering the main road from the ramp must accelerate to a certain speed on the acceleration lane and then find an appropriate time to merge into a target gap on the main road. However, due to drivers' inaccurate estimation of the size of the target merging gap or speed differences between ramp and main road vehicles, it is common for merging vehicles to have difficulty smoothly merging into the gap or to be involved in rear-end collisions or side-swipes with main road vehicles shortly after changing lanes. According to data from relevant authorities, the number of traffic accidents at highway entrance ramps is four times higher than on regular road sections. In comparison to regular road sections, the merging of main road traffic and ramp traffic in the highway entrance ramp area significantly reduces the area's traffic capacity.

#### *2) Analysis of traffic efficiency in highway ramp merge areas:*

When highway ramp vehicles attempt to merge onto the mainline, drivers of mainline vehicles are highly likely to decelerate or take other actions to yield. The acceleration of ramp vehicles merging into the mainline disrupts the flow of traffic, especially during peak traffic periods when traffic volume is high. As a result, the ramp merge point often becomes a bottleneck area in traffic flow, leading to traffic delays and congestion, turning the highway ramp merge area into a bottleneck for the entire highway. When the traffic volume upstream of the merge area increases and even exceeds the maximum capacity of the roadway segment, ramp vehicles may be unable to merge onto the mainline. Vehicles will queue up on the ramp and gradually extend onto the mainline. When the highway ramp merge area is congested, both the maximum capacity and the traffic dissipation rate significantly impact travel speed. Firstly, as traffic approaches or exceeds the maximum capacity of the roadway, any minor disturbance can trigger traffic instability, requiring vehicles to decelerate to maintain a safe distance, directly reducing travel speed. Secondly, in congested conditions, the traffic dissipation rate becomes a critical factor, determining the speed at which traffic returns to a normal state. If the dissipation rate is low, even after the initial congestion factors have been resolved, the recovery of traffic flow will be slow, further impacting travel speed. Additionally, the "wave effect" in high-density traffic flow and cautious and hesitant behavior of drivers in congestion increase the gap time between vehicles, reducing overall traffic efficiency.

### B. *The Judgment Process for Coordinated Control*

No coordinated control refers to the unrestricted free movement of both mainline and ramp vehicles. In this paper, the free movement state of vehicles is defined as follows: mainline vehicles travel at a constant speed of $v_0$, while ramp vehicles travel at a constant speed of $v_{R0}$ on the ramp until the end of the ramp, then accelerate at a rate of $a_r$ m/s$^2$ on the acceleration lane to merge directly into the mainline at the same speed as the mainline vehicles.

In this state of free movement, two main scenarios are likely to occur:

(1) When ramp vehicles can safely merge into the mainline, there is no conflict between ramp vehicles and mainline vehicles, thus no adjustment of vehicles is required. When vehicles can merge into the mainline safely without coordination, it incurs the minimum cost, as no vehicle needs to be adjusted.

(2) When conflicts arise between ramp vehicles and mainline vehicles, coordinated control of both ramp and mainline vehicles is necessary. This paper proposes two methods of coordinated control: mainline priority and ramp priority coordinated control methods. Through these two coordinated control methods, ramp vehicles can safely merge into the mainline. Both of these coordinated control methods incur certain costs when ensuring the safe merging of ramp vehicles into the mainline. These costs may involve sacrificing vehicle travel speed or increasing vehicle fuel consumption.

The process involves determining whether the ramp vehicle can merge into the mainline without coordinated control and conflict with mainline vehicles. If there is no conflict, the ramp vehicle merges safely into the mainline, achieving successful merging, and the process ends. If there is a potential conflict, coordinated control is applied to the relevant mainline and ramp vehicles. Coordinated control methods include mainline priority and ramp priority. Ultimately, with the assistance of coordinated control, the ramp vehicle merges safely into the mainline, achieving successful merging, and the process ends.

*C. The Basic Idea of Mainline Priority is to Prioritize the Flow of Mainline Traffic*

The application of vehicle-road deep coordination technology can significantly optimize various challenges in highway ramp merging areas, significantly improving the safety and operational efficiency of the transportation system. This paper proposes a vehicle-road deep coordination method with mainline priority, which requires precise positioning of mainline vehicles that may conflict with ramp vehicles. Then, it searches for the minimum merge gap that allows the ramp vehicle to merge safely near the conflicted mainline vehicle. The ramp vehicle adjusts its speed to reach this gap and merges safely into the mainline. The Road Section Management Units plans the vehicle trajectories in advance based on the above model and then transmits them to the autonomous vehicles, which follow the pre-planned trajectories.The process diagram for ramp vehicles merging into the mainline with mainline priority in a vehicle-road deep coordination environment is shown in Fig. 1

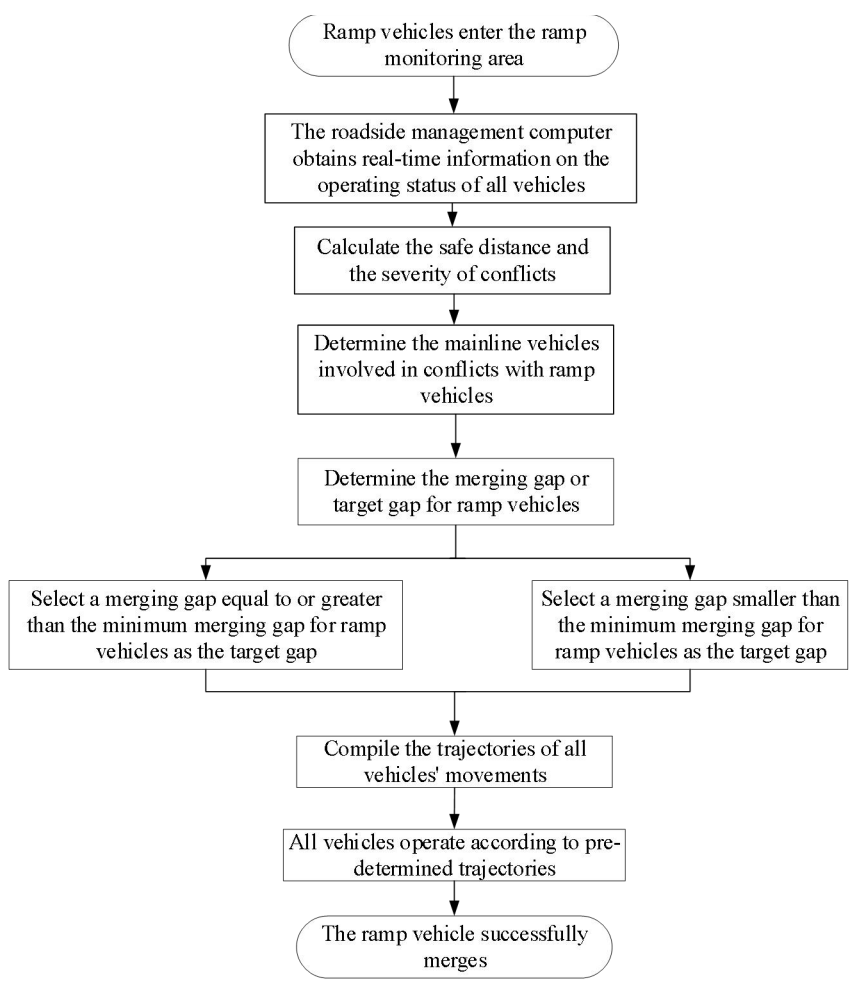

Fig. 1. The process diagram for ramp vehicles merging into the mainline based on mainline priority

In the condition of free movement, it is easy for ramp vehicles to conflict with mainline vehicles when merging into the mainline in highway ramp merging areas. Therefore, the key to allowing ramp vehicles to merge safely into the mainline is to identify and resolve conflicts between ramp and mainline vehicles. The first step in resolving conflicts between ramp and mainline vehicles is to determine which mainline vehicle will conflict with the ramp vehicle. Then, based on the size of the gap around the conflicted mainline vehicle, the target gap for the ramp vehicle to merge into is selected. Initially, gaps around conflicted mainline vehicles that are equal to or larger than the minimum merge gap for ramp vehicles are selected as target gaps. However, if selecting such a gap does not satisfy the acceleration and merging point requirements for the ramp vehicle, it will not be chosen as the target gap. If there are no gaps around conflicted mainline vehicles that meet the requirements, gaps smaller than the minimum merge gap for ramp vehicles will be selected as target gaps. If such gaps are selected, mainline vehicles in front and behind the target gap need to adjust their speeds during the journey to leave a gap equal to or larger than the minimum merge gap for ramp vehicles. After selecting the target gap, the acceleration for the ramp vehicle to merge into the target gap and the acceleration and time for mainline vehicles to adjust their speeds are calculated. Then, based on the calculated data, the trajectory for adjusting vehicles,the functional relationship between the station number and time for vehicles, is prepared. Vehicles that are not adjusted continue to travel at their previous speeds. After the Road Section Management Units prepares the vehicle trajectories, they are sent to the respective vehicles, and all vehicles follow the pre-planned trajectories. Then, ramp vehicles can merge safely.

First, select a gap before or after the mainline vehicle conflicting with the ramp that is greater than or equal to the minimum acceptable merging gap as the target merging gap for the ramp vehicle to merge into.

Next, plot the spatiotemporal diagram depicting the changes in milepost over time for 7 mainline vehicles and 1 ramp vehicle, as shown in Fig. 2, along with a zoomed-in view in Fig. 3. The mainline vehicles are all traveling at a speed of 100 km/h, while the ramp vehicle travels at 60 km/h on the ramp and accelerates at 2 m/s² on the acceleration lane until merging directly into the mainline when its speed matches that of the mainline vehicles.

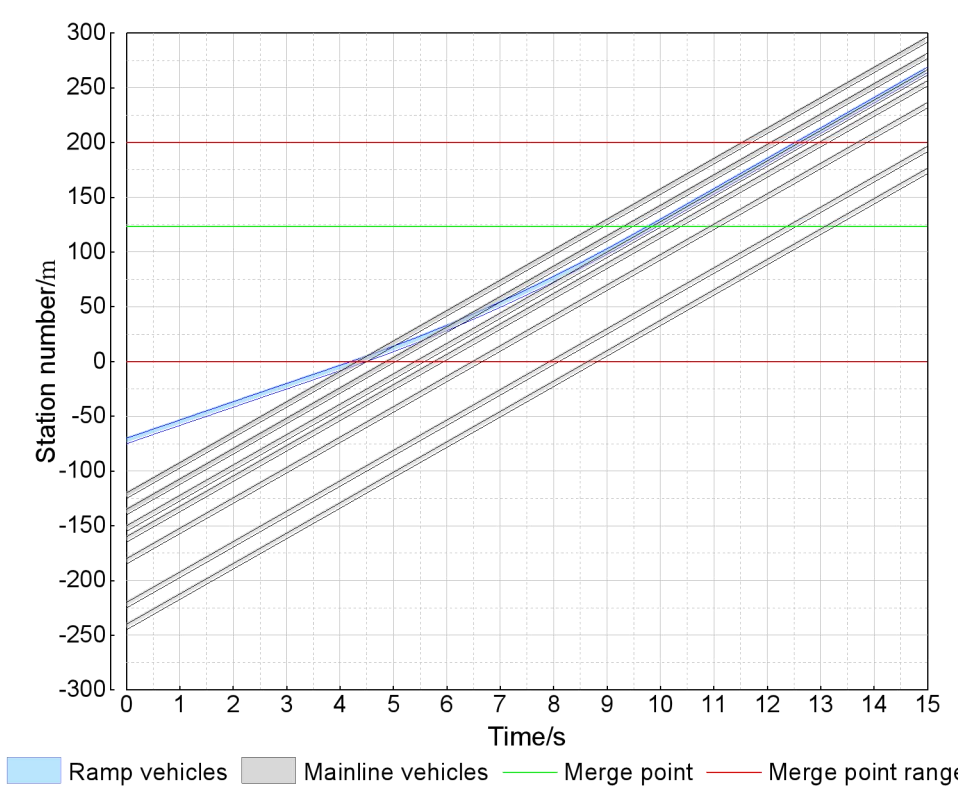

Fig. 2. The spatiotemporal diagram before adjustment.

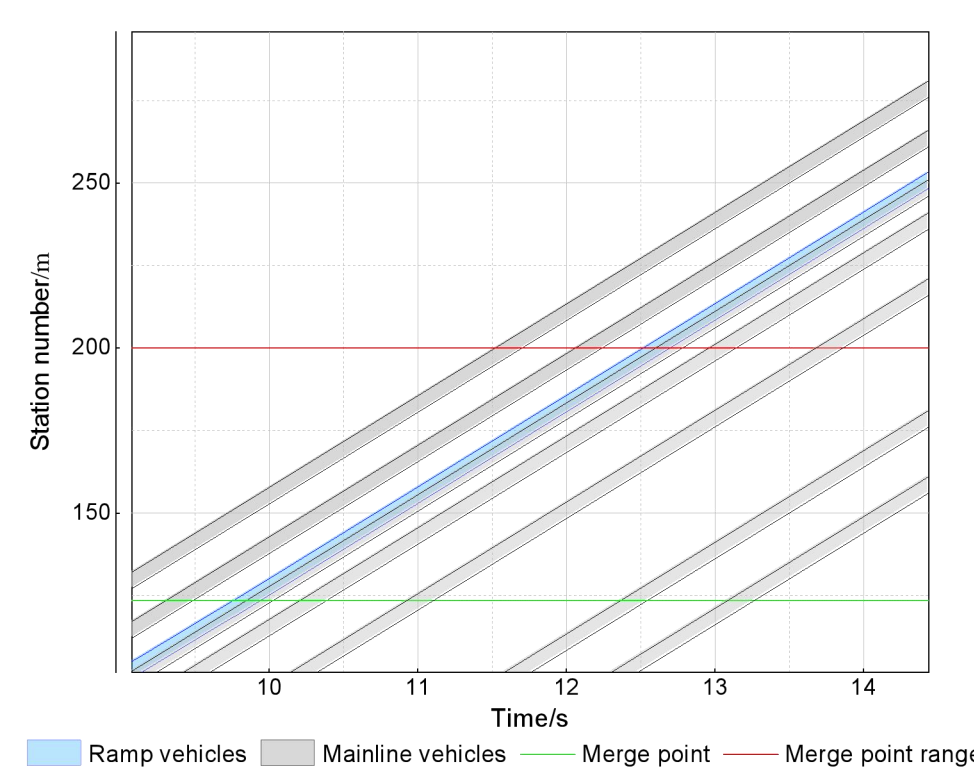

Fig. 3. The enlarged view of the spatiotemporal diagram before adjustment.

Using the aforementioned method, we can calculate the acceleration of the ramp vehicle and the merging point after implementing cooperative methods. We then adjust the positions of vehicles that would have collided without cooperative control. The resulting spatiotemporal diagram of the vehicles, along with localized enlargements (Fig. 4 and Fig. 5, respectively).

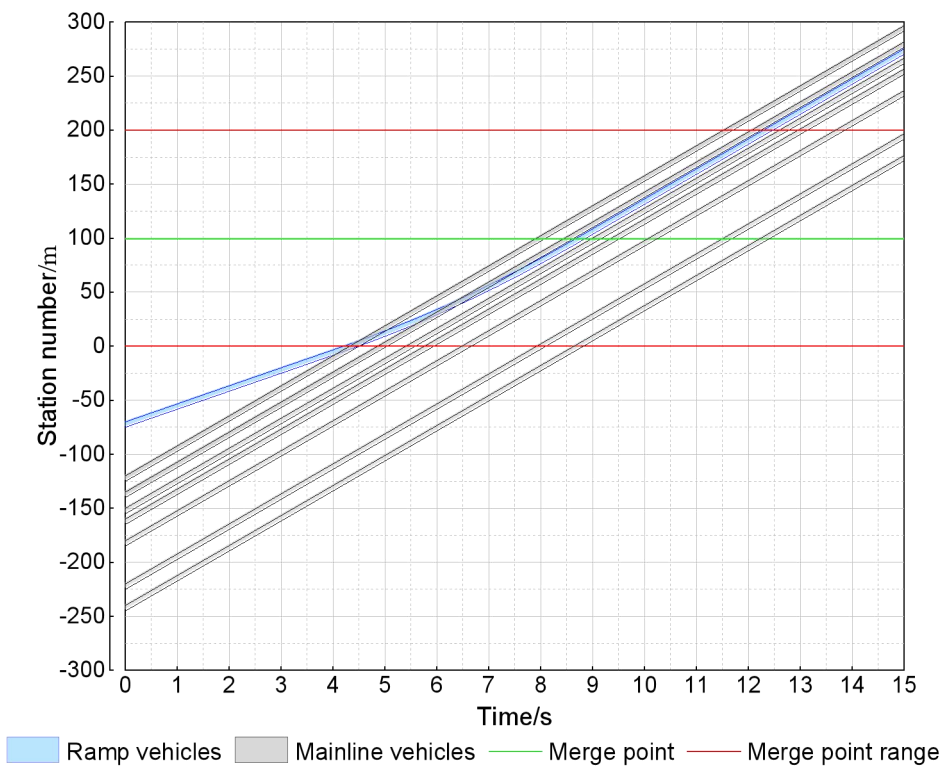

Fig. 4. The adjusted spatiotemporal diagram of the vehicles.

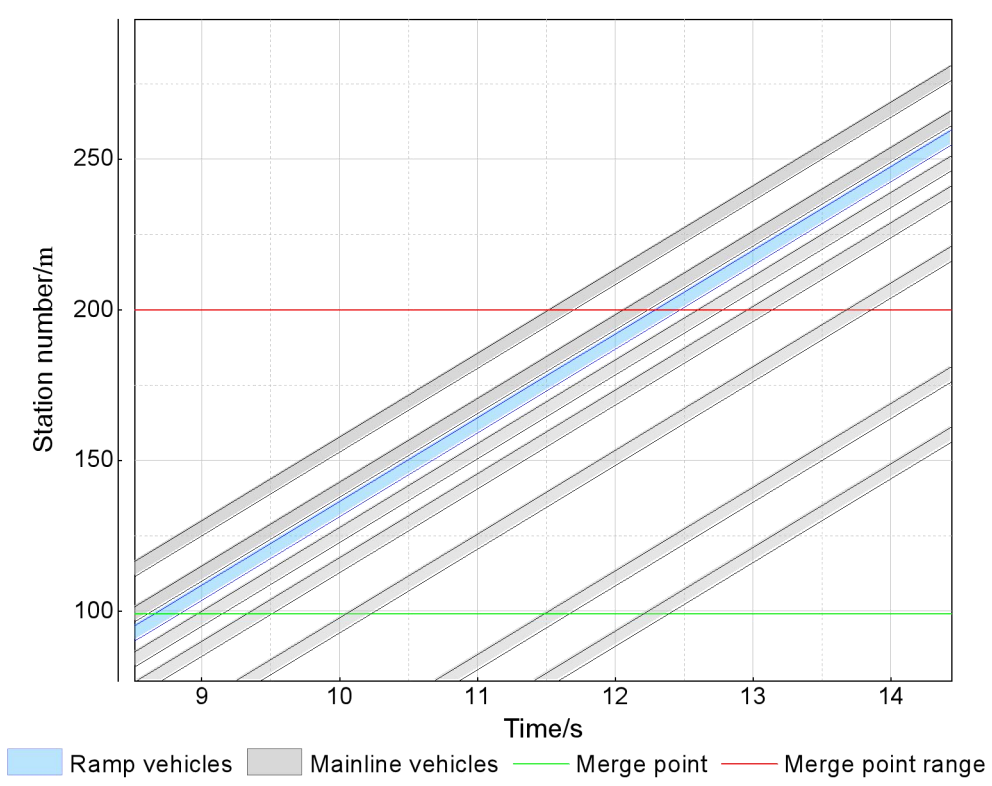

Fig. 5. The localized enlargement of the adjusted spatiotemporal diagram of the vehicles.

## D. *Basic Idea of Ramp Prioritization*

The key to ramp vehicles merging into the mainline in the merging area of a highway ramp is to find suitable merging gaps. In the mainline priority strategy, the focus is on identifying target gaps on the mainline, allowing ramp vehicles to find opportunities to merge into the mainline. On the other hand, in the ramp priority strategy, the merging position of ramp vehicles is determined, allowing the vehicles before and after the target gap on the mainline to move freely or adjusting the mainline vehicles to enable ramp vehicles to reach a specified position. Subsequently, the ramp vehicles can smoothly merge into the mainline.

# V. Simulation and Evaluation

Average delay is a crucial traffic metric, especially when evaluating the effectiveness of highway ramp merging, as it provides key insights into traffic flow and efficiency. It reflects traffic congestion, merging efficiency, traffic safety, and the effectiveness of cooperative control.

In this paper, simulations were conducted with mainline traffic volumes of 800 veh/h/lane, 1200 veh/h/lane, and 1800 veh/h/lane, and ramp traffic volumes of 200 veh/h/lane, 300 veh/h/lane, and 500 veh/h/lane. Three strategies were simulated: mainline priority, ramp priority, and SUMO's own cooperative control. This resulted in nine different traffic flow scenarios with corresponding average delay times, as shown in Fig. 6.

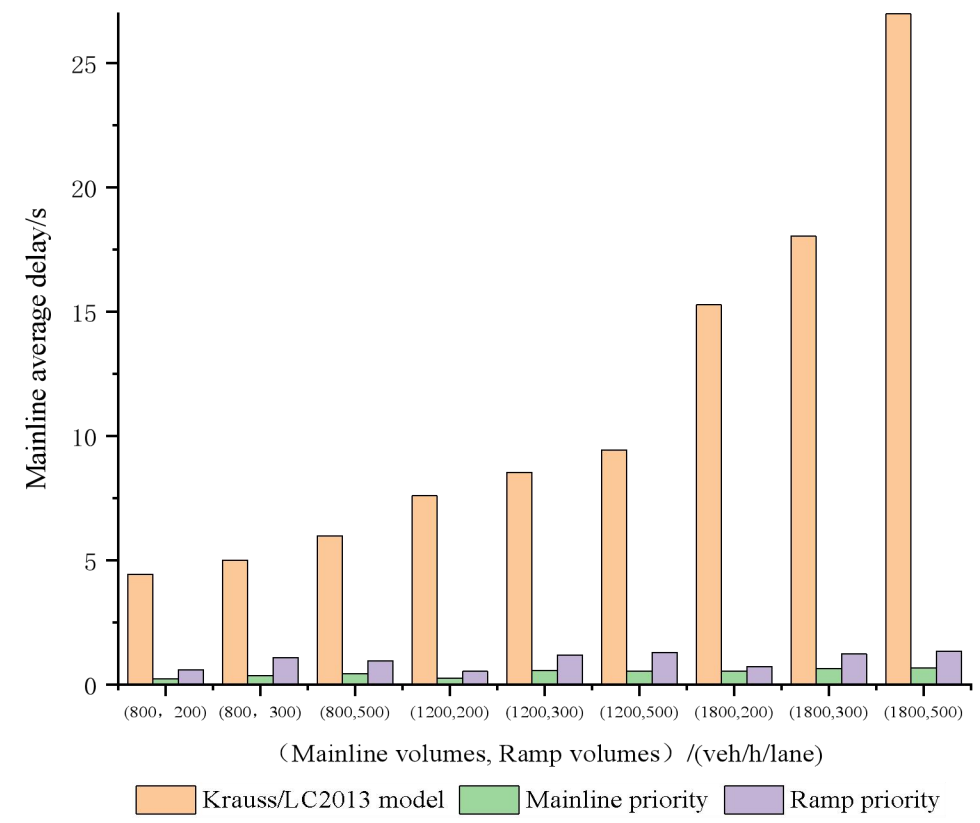

Fig. 6. Mainline average delay graph

From Fig. 6. it can be observed that among the 9 different traffic flow conditions, the use of the KRAUSS/LC2013 model results in the highest average delay on the mainline, indicating that this method may lead to a decrease in the efficiency of mainline traffic. On the other hand, while the ramp priority strategy reduces delays on the ramp, it does so at the expense of mainline smoothness, potentially causing frequent deceleration of mainline vehicles. In contrast, the mainline priority control strategy exhibits the lowest average delay, implying that it not only maintains the smoothness of mainline traffic but also effectively manages the merging of ramp vehicles, achieving a good balance between merging efficiency and safety, thus enhancing both traffic flow efficiency and driving safety. These factors collectively demonstrate the superior comprehensive performance of the mainline priority coordination control strategy.

The ramp average delay reflects:Fig. 7. shows the ramp delay graphs under the three coordinated control strategies ..

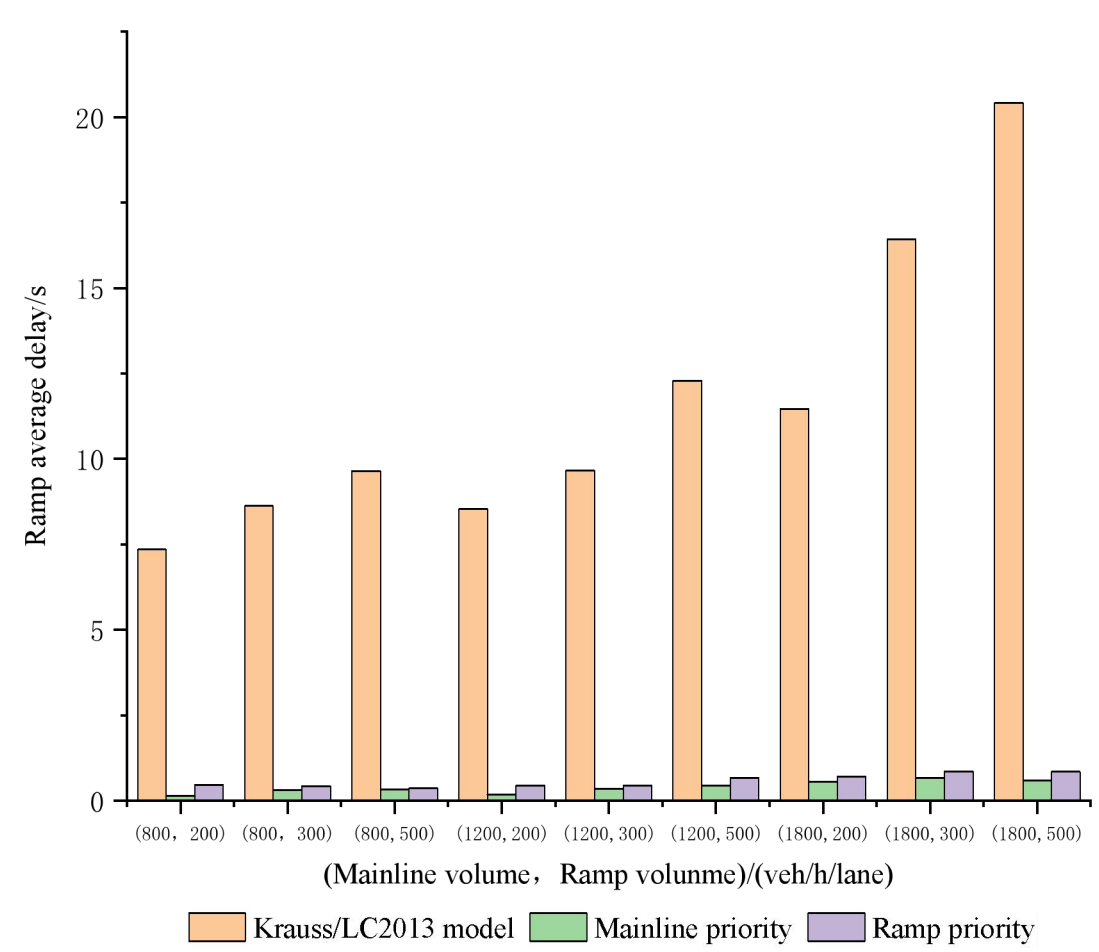

Fig. 7. Ramp average delay graph

From Fig.7 . it can be observed that among the 9 different traffic flow conditions, the ramp average delay is highest under the KRAUSS/LC2013 model, indicating that this method may have failed to effectively address the merging demands on the ramp, resulting in high delays on the ramp. In contrast, while the ramp priority strategy reduces ramp delays, it does not achieve the same level of effectiveness as the mainline priority strategy, which achieves smaller ramp average delays. This indicates that among these three coordinated controls, the mainline priority strategy allows ramp vehicles to smoothly and rapidly enter the mainline, thus being the most efficient. This not only improves merging efficiency and reduces the likelihood of drivers engaging in aggressive behaviors due to delays but also likely enhances traffic safety while improving road utilization efficiency. Therefore, the mainline priority control strategy performs the best among the three coordinated control strategies.

Through simulation analysis, the information sharing mechanism demonstrates excellent performance in terms of timeliness, stability, and security, with stable and secure data transmission across various packet sizes. The real-time sharing mechanism provides drivers with timely and comprehensive driving information, optimizing driving decisions.

## VI. CONCLUSIONS

This paper, based on deep vehicle-road coordination, designs a real-time information sharing mechanism among vehicles in neighboring areas and studies the construction method of spatiotemporal global views of highway traffic infrastructure operations. Furthermore, to proactively eliminate conflicts between vehicles and enhance road efficiency while ensuring safety, this paper proposes a technical solution through the adoption of a method where the Road Section Management Unit pre-plans vehicle spatiotemporal trajectories and has them executed by Vehicle Intelligent Units, thus achieving spatiotemporal coordination to preemptively eliminate conflicts between vehicles. It also derives a method to calculate safe distances between vehicles in deep vehicle-road coordination scenarios, quantifies the urgency of vehicle collisions based on collision acceleration, and proposes two spatiotemporal coordinated control methods: mainline priority and merge lane priority.Parameters such as vehicle output power, road gradient, and road surface friction coefficient were not considered in this book. Future research could explore the impact of these factors and further refine methods for pre-planning vehicle spatiotemporal trajectories.